 \documentclass[final,5p,times,twocolumn]{elsarticle}

\usepackage{enumerate}
\usepackage{amsmath}
\usepackage{amsfonts}
\usepackage{amssymb}
\usepackage[utf8]{inputenc}
\usepackage[T1]{fontenc}
\usepackage{mathtools}
\usepackage{wasysym}
\usepackage{accents}
\usepackage[svgnames]{xcolor}
\usepackage{url}
\usepackage[colorlinks=true,citecolor=blue]{hyperref}
\usepackage{url}
\usepackage{graphicx}
\usepackage{bm}
\usepackage{tikz}
\usepackage{xcolor}
\usetikzlibrary{arrows,decorations.markings}
\usepackage{csquotes}

\usepackage[skins,theorems,most]{tcolorbox}
\tcbset{highlight math style={left=3mm,right=3mm,top=3mm,bottom=3mm,enhanced, colframe=black!60!black,colback=white,arc=4pt,boxrule=1pt}} 
\usepackage{orcidlink}

\definecolor{mygreen}{rgb}{0,0.7,0}

\usepackage[normalem]{ulem}

\usepackage{amssymb}
\usepackage{lipsum}

\journal{Physics Letters B}

\begin{document}
\begin{frontmatter}

\author[JP]{Pedro H. Morais\,\orcidlink{0000-0002-2226-3579}}
\ead{phm@academico.ufpb.br}
\author[JP2,JP3]{Iarley P. Lobo\,\orcidlink{0000-0002-1055-407X}}
\ead{lobofisica@gmail.com}
\author[CP]{Christian Pfeifer\,\orcidlink{0000-0002-1712-6860}}
\ead{christian.pfeifer@ut.ee}
\author[RA]{Rafael {Alves Batista}\,\orcidlink{0000-0003-2656-064X}}
\ead{rafael.alvesbatista@uam.es}
\author[JP]{Valdir B. Bezerra\,\orcidlink{0000-0001-7893-0265}}
\ead{valdir@fisica.ufpb.br}

\affiliation[JP]{organization={Department of Physics, Federal University of Para\'iba},
            city={João Pessoa},
            postcode={58059-900}, 
            state={PB},
            country={Brazil}}
\affiliation[JP2]{organization={Department of Chemistry and Physics, Federal University of Para\'iba},
            addressline={Rodovia BR 079 - Km 12}, 
            city={Areia},
            postcode={58397-000}, 
            state={PB},
            country={Brazil}}
\affiliation[JP3]{organization={Physics Department, Federal University of Lavras},
            city={Lavras},
            postcode={37200-000}, 
            state={MG},
            country={Brazil}}
\affiliation[CP]{organization={ZARM, University of Bremen},
            city={Bremen},
            postcode={28359},
            country={Germany}}

\affiliation[RA]{organization={Instituto de F\'isica Te\'orica UAM-CSIC},
            addressline={C/ Nicol\'as Cabrera 13-15}, 
            city={Madrid},
            postcode={28049},
            country={Spain}}

\title{Modified particle lifetimes as a signature of deformed relativity}
           
\begin{abstract}
We demonstrate a compatibility between the relativity principle and the clock postulate in deformed special relativity, by identifying the relevant deformed Lorentz transformations in position space between arbitrary frames. This result leads to a first-principles correction to the dilated lifetime of fundamental particles. It turns out that these modified time dilations offer a way to scrutinize Lorentz invariance (or deviations thereof) to high precision.
\end{abstract}




\end{frontmatter}




\section{Introduction}
The characterization of what should be a quantum spacetime is one of the routes that may lead us to an appropriate description of quantum gravity. We expect that such a challenging task should pass through intermediary steps before its full realization from the theoretical and experimental points of view. For this reason, it is plausible to expect that corrections to the Riemannian and general relativistic descriptions of gravity should become manifest once we advance toward the Quantum Gravity (QG) scale. In this sense, quantum gravity phenomenology plays a fundamental role by translating the intuition of this area to observables, recognizable in our current treatment of spacetime physics (for reviews on the subject we refer the reader to \cite{Amelino-Camelia:2008aez,Addazi:2021xuf}).
\par
It is known that there exist formulations of such effective quantum gravity spacetime geometries, in which some of the cornerstones of modern physics, like the physical equivalence between local inertial observers, are preserved \cite{Amelino-Camelia:2000stu}. Those formulations that incorporate an invariant length/energy scale, which is expected to exist from various fundamental quantum gravity models, in a relativistic way, are known as Deformed Special Relativity (DSR) models \cite{Magueijo:2002am,Majid:1994cy,Bruno:2001mw,Barcaroli:2015xda,Girelli:2006fw}. They extend Einstein's first postulate and their impact on observables is intensively studied. Among the possibilities for the realization of this idea, Finsler geometry stands out when one searches for a continuous description of the kinematics of a particle in a curved spacetime with a fundamental length \cite{Girelli:2006fw,Lobo:2020qoa,Lobo:2016xzq,Zhu:2023mps}. It describes the spacetime geometry fully in terms of an arc-length functional (for the usefulness of different commutative geometries for the description of phase and configuration spaces, please refer to the reviews \cite{Pfeifer:2019wus,Albuquerque:2023icp,Zhu:2023kjx}). Finslerian geodesics turn out to be the deformed trajectories of massless particles that are analyzed in the rich phenomenology of time delays from gamma-ray bursts \cite{Zhu:2022blp} and isometries of the Finsler metric are connected to local deformations of the Lorentz symmetry \cite{Amelino-Camelia:2014rga}. This means for example, that if we call $E_{\text{Pl}}$ the Planck energy, then the kinematics of a particle of mass $m$, energy $E$ and momentum $|\vec{p}|=|p|$, subject to a modified dispersion relation (MDR) of the form ($\eta^{(n)}$ is a dimensionless parameter that controls the perturbative approach that we are going to follow and $n$ is the leading order of the perturbation of the special relativistic dispersion relation)\footnote{We are using natural units such that $c=\hbar=1$.}
\begin{equation}
 E^2-p^2=m^2+\eta^{(n)}\frac{|p|^{n+2}}{E_{\text{Pl}}^n}\,\label{mdr}
\end{equation}
are determined by the arc-length functional,
\begin{equation}
    s=\int F(\dot{t},\dot{x})d\lambda\, ,
\end{equation}
where 
\begin{equation}
    F(\dot{t},\dot{x})=\sqrt{\dot{t}^2 - \dot{x}^2}+\frac{\eta^{(n)}}{2}\left(\frac{m}{E_{\text{Pl}}}\right)^n\frac{|\dot{x}|^{n+2}}{(\dot{t}^2 - \dot{x}^2)^{\frac{n+1}{2}}}\, . \label{finsler-function} 
\end{equation}
Here, ``dot'' means derivative with respect to the parameter $\lambda$ and $|\dot x| = |\dot{\vec{x}}|$. 
For simplicity, we shall assume motion in one spatial dimension, $x$, in the following. The function $F$ is called Finsler function and this result has been derived in more details and generality in \cite{Lobo:2020qoa}. This equivalence of treatments is due to the fact that a Finsler function above is the Lagrangian derived from a Legendre transformation \cite{Rodrigues:2022mfj} of the nonquadratic Hamiltonian defined by the MDR \eqref{mdr}, as was originally discussed in~\cite{Girelli:2006fw}.

\section{Time dilation from the Clock Postulate}
Recently, a novel aspect of the Finslerian description of quantum spacetime was found in \cite{Lobo:2020qoa}. By applying the Clock Postulate (CP) (which states that the proper time an observer measures between two events is given by the arc-length of its worldline in spacetime between the two events), we found that the dilated laboratory lifetime $t$ of a particle with velocity $v=|d\vec{x}/dt|$ relative to the laboratory (lab) and proper rest frame lifetime $\tau$ is 
\begin{equation}
    t = \gamma \tau\left[1-\frac{\eta^{(n)}}{2}\left(\frac{m}{E_{\text{Pl}}}\right)^n(\gamma^2-1)^{\frac{n+2}{2}}\right]\, ,\label{rest-isometry}
\end{equation}
where $\gamma^{-1}=\sqrt{1-v^2}$. We clearly see the Planck-scale corrections beyond Special Relativity (SR) induced by the MDR \eqref{mdr}. In order to connect this expression with observations, it is necessary to express the velocity $v$ in terms of the particle's energy in Finsler geometry, which simply reads
\begin{align}
\begin{split}
    E&=m\frac{\partial F}{\partial \dot{t}}\Bigg{|}_{\lambda=t}\\
    &=m\gamma\left[1 - \frac{\eta^{(n)}}{2}(n+1)\left(\frac{m}{E_{\text{Pl}}}\right)^n(\gamma^2-1)^{\frac{n+2}{2}}\right].
    \label{gamma-energy}
\end{split}
\end{align}

Curiously, as we discussed in \cite{Lobo:2021yem}, this expression is actually a deformed Lorentz transformation from the rest frame to the laboratory. Expression \eqref{gamma-energy} can be inverted, giving 
\begin{equation}
    \gamma=\frac{E}{m}\left[1 +\frac{\eta^{(n)}}{2}(n+1)\left(\frac{m} {E_{\text{Pl}}}\right)^n\left[\left(\frac{E}{m}\right)^2-1\right]^{\frac{n+2}{2}}\right]\, .
\end{equation}
\noindent
From this expression, we easily find from the CP the Finslerian description of the dilated lifetime of a particle. Let $\tau$ be the particle's lifetime at rest and $m$ and $E$ be it's mass and energy which obey a MDR of the form \eqref{mdr}. Then, the particle lifetime $t$ in the laboratory frame is
\begin{equation}
\tcbhighmath{ t=\gamma_{\text{CP}}\tau=\frac{E}{m}\left[1+\frac{n\eta^{(n)}}{2}\left(\frac{|p|}{m}\right)^{2}\left(\frac{|p|}{E_{\text{Pl}}}\right)^n\right]\tau\, .}\label{cp-lifetime}
\end{equation}
\par
We call the modified Lorentz factor that dilates the lifetime $\gamma = \gamma_{\text{CP}}$, since it was calculated from an extension of the clock postulate to an effective quantum spacetime described in terms of Finsler geometry that is used for describing the kinematics of a particle subject to a MDR.
\par
A similar effect was described preliminarily in \cite{Trimarelli:2022wdd}, in which a corrected Lorentz factor is suggested to be $\gamma_{\text{LIV}}=E/m_{\text{LIV}}$, where $m_{\text{LIV}}^2=m^2+\eta^{(n)}|p|^{n+2}/E_{\text{Pl}}^n$ is the right-hand side of Eq.\eqref{mdr} (LIV stands for Lorentz Invariance Violation). The first order (and dominant) correction of this expression gives
\begin{equation}
\gamma_{\text{LIV}}=E/m_{\text{LIV}}\approx \frac{E}{m}\left[1-\frac{\eta^{(n)}}{2}\left(\frac{|p|}{m}\right)^2\left(\frac{|p|}{E_{\text{Pl}}}\right)^n\right]\, .\label{gamma-liv}
\end{equation}

The expressions for $\gamma_{\text{LIV}}$ and $\gamma_{\text{CP}}$ look similar at first order (one simply translates superluminal effects from $\gamma_{\text{LIV}}$ to be subluminal in $\gamma_{\text{CP}}$). However, only in the CP case the concept of time emerges in a natural way due to the Finslerian approach employed. In the LIV case one could say that there seems to be no deeper reason to suppose that a Lorentz factor that dilates lifetimes should be modified the way it is proposed. 
\par
Interestingly, in both cases, this kind of correction presents an amplifying factor given by $(|p|/m)^2$, which can furnish large values for ultra-high-energy cosmic rays (UHECRs). This is the reason why dilated lifetimes have been considered as potential observables in the search for quantum gravity and deviations from Lorentz invariance \cite{Trimarelli:2022wdd,Cowsik:2012qm}, in addition to other effects such as modified interaction thresholds~\cite{PierreAuger:2021tog}.
\par
Despite the clear physical interpretation and mathematical formulation of the CP approach, one could be tempted to state that an actual comparison between times in different frames {\it must} be derived from an actual map between observers. In order to be coherent with the DSR roots of the Finsler relation with quantum gravity phenomenology, such an effect must come from a Deformed Lorentz Transformation (DLT) involving spacetime coordinates -- a step that is missing so far in this approach. 
\par
This is precisely the goal of this letter. We seek therefore to show that in this DSR scenario, just like in SR, the result concerning the CP is actually an isometry of the Finsler measure or a DLT between frames that move relative to each other with velocity $v$.

\section{Compatibility between the Clock Postulate and a Deformed Lorentz Transformation}
To prove that Eq.\eqref{cp-lifetime} is indeed a DLT, we use the geodesic equation that determines the relation $x(t)$. To find it, we use a conserved quantity given by the spatial momentum $p=m\partial F/\partial\dot{x}$, as this expression is parametrization-invariant, we use the laboratory time as a parameter and solve this equation for $dx/dt$. Finally, we use Eq.\eqref{cp-lifetime} to express this solution as a function of $\tau$, $E$, $p$ and $m$ as\footnote{In fact, the relation between a propagation distance $L_{xy}=|x|$, the transverse momentum $|p|=p_{\text{T}}$, the mass of a particle from the PDG $m=M_{\text{PDG}}$ and the proper time $\tau$ is the basis for the measurement of particles' proper lifetimes in accelerators \cite{ALICE:2023ecf}. Therefore, our result describes discrepancies that could emerge for measurements done with future experiments (with higher energies than those attainable today) and with better precision \cite{Lobo:2023yvi}.}
\begin{equation}
   \tcbhighmath{x=-\frac{p\, \tau}{m}\left[1+\frac{\eta^{(n)}}{2}\frac{(2m^2+nE^2)}{m^2}\left(\frac{|p|}{E_{\text{Pl}}}\right)^n\right]\, .}\label{cp-x}
\end{equation}

This is basically the geodesic solution in the proper time parametrization (the solution for the other cartesian spatial coordinates is trivial as we are considering one-dimensional motion). If we use the above expression along with \eqref{cp-lifetime} to calculate the Finsler function \eqref{finsler-function}, a direct calculation shows that for on-shell particles
\begin{align}
F(\dot{t},\dot{x})&=\sqrt{\dot{t}^2 - \dot{x}^2}+\frac{\eta^{(n)}}{2}\left(\frac{m}{E_{\text{Pl}}}\right)^n\frac{|\dot{x}|^{n+2}}{(\dot{t}^2 - \dot{x}^2)^{\frac{n+1}{2}}}\nonumber\\
&=\dot{\tau}=F(\dot{\tau},0)\, .\label{isometry-cond-rest}
\end{align}

{\it This proves that the set of transformations given by Eqs.\eqref{cp-lifetime} and \eqref{cp-x} corresponds to an isometry in Finsler geometry, therefore, they constitute a Deformed Lorentz Transformation.}
\par
An alternative expression for this transformation can be found by expressing $E$ and $p$ as a function of the velocity $v$ from $p_{\mu}=m\partial F/\partial \dot{x}^{\mu}$ in the lab time parametrization. In this case, the transformations for $t$ and $x$ are simply
\begin{align}
     t &= \gamma\tau \left[1-\frac{\eta^{(n)}}{2}\left(\frac{m}{E_{\text{Pl}}}\right)^n(\gamma^2-1)^{\frac{n+2}{2}}\right]=\gamma_{\text{CP}}\tau\, ,\label{transf1}\\
    x &= v\gamma\tau \left[1-\frac{\eta^{(n)}}{2}\left(\frac{m}{E_{\text{Pl}}}\right)^n(\gamma^2-1)^{\frac{n+2}{2}}\right]=v \gamma_{\text{CP}}\tau\, .
    \label{transf2}
\end{align}

It is straightforward to verify that indeed $v$ is the velocity of the particle in the lab frame, since we check that $dx/dt=v$.
\par
This is a remarkable result. For the first time, we have simultaneously a DLT involving space and time coordinates and the boost parameter is undoubtedly identified as the velocity of a particle $dx/dt$. Besides that, this result is compatible with the CP (just like in SR), which allows us to indeed describe what would be a Planck-scale correction to the twin paradox. For this reason, we are confident to state that Eq.\eqref{cp-lifetime} actually defines a DSR Lorentz factor
\begin{equation}
\tcbhighmath{ \gamma_{\text{CP}}=\gamma_{\text{DSR}}\, .}\label{cp-dsr}
\end{equation}

Let us compare our findings here with the ones made in \cite{Trimarelli:2022wdd} for the LIV case, which are currently being analyzed using UHECRs. One may be tempted to translate the results found in that paper using $\eta^{(n)}\mapsto -n\eta^{(n)}$ (since only very-high energy relativistic particles would effectively contribute to the effect). But, we should notice that besides the modification in the particle's lifetime, also a modified velocity is used as input
\begin{equation}
v_{\text{LIV}}=\beta_{\text{LIV}}=\frac{|p|}{m\gamma_{\text{LIV}}}\approx \frac{|p|}{E}\left[1+\frac{\eta^{(n)}}{2}\left(\frac{|p|}{m}\right)^2\left(\frac{|p|}{E_\text{Pl}}\right)^n\right]\, .\label{vliv}
\end{equation}

In our case, the relation between the velocity $v$ and the momenta is naturally given by the definition of the spatial physical momentum
\begin{equation}\label{eq:p}
    p=m\frac{\partial F}{\partial \dot{x}}\Bigg{|}_{\lambda=t}=-mv\gamma\left[1 + \frac{\eta^{(n)}}{2}\frac{(mv)^n(v^2-2-n)}{E_{\text{Pl}}^n (1-v^2)^{\frac{n+2}{2}}}\right]\, ,
\end{equation}
from which we can calculate its absolute value $|p|$ and, using the relation between the Lorentz factor and the energy \eqref{gamma-energy}, we derive the following
\begin{equation}
\tcbhighmath{ v_{\text{DSR}}=\beta_{\text{DSR}}=\frac{|p|}{E}\left[1+\frac{(2+n)\eta^{(n)}}{2}\left(\frac{|p|}{E_{\text{Pl}}}\right)^n\right]\, .}\label{vdsr}
\end{equation}

As expected, this result is actually the velocity of the particle defined from the MDR \eqref{mdr}, since
\begin{equation}
    v=\Bigg{|}\frac{\partial E}{\partial p}\Bigg{|}\stackrel{\text{MDR}}{=\mathrel{\mkern-3mu}=\mathrel{\mkern-3mu}=}v_{\text{DSR}}\, .\label{v-vdsr}
\end{equation}
For this reason, even in a LIV scenario, one should use Eq.\eqref{vdsr} instead of \eqref{vliv}. So, we can actually drop the symbol DSR out of the velocity in \eqref{vdsr}, since it is simply the velocity of the particle read from the MDR. With this observation, we complete the analysis connecting the rest and the lab frames. The next natural step consists in connecting general spacetime rectangular coordinates of different frames that move relative to each other with velocity $v$, i.e., $(t,x)\mapsto (t',x')$, which reduces to the previous case when the target frame is $(\tau,0)\mapsto (t,x)$.

\section{General Deformed Lorentz Transformation}
In order to generalize the previous isometry such that we connect two arbitrary rectangular coordinates, it is sufficient to propose a transformation involving the boost parameter $v$, the spacetime coordinates $(t,x)$ and the velocities $(\dot{t},\dot{x})$ that reduces to the previous case when the target frame obeys $x=0=\dot{x}$. Since this is a transformation that shall leave the Finsler function invariant (and consequently the metric), it should not depend on the parametrization $\lambda$ of the velocities $(\dot{t},\dot{x})$.

The functions that naturally satisfy this requirement are the momentum components
\begin{align}
    E(\dot t, \dot x) = m \frac{\partial F(\dot t, \dot x)}{\partial \dot t},\quad 
    p(\dot t, \dot x) = m \frac{\partial F(\dot t, \dot x)}{\partial \dot x}\,,
\end{align}
which satisfy $E=m$ in the particle rest frame, i.e. for $x=0=v$, see \eqref{gamma-energy} and \eqref{eq:p}.

A generalisation of the transformations \eqref{transf1} and \eqref{transf2} for transformations between arbitrary frames $(t,x)\mapsto (t',x')$ is constructed from combinations of factors of the type $E^{n-r}|p|^r/E_{\text{Pl}}^n$, where $E$ and $p$ are treated as functions of $(\dot{t},\dot{x})$. A general ansatz for such transformations to first order in $\eta^{(n)}$ is of the following form:

    \begin{align}
        t' = &\,\,\, (t+xv)\gamma +\frac{\eta^{(n)}}{2}t\gamma \left(\frac{E}{E_{\text{Pl}}}\right)^n\left(\gamma ^2-1\right)^{\frac{n+2}{2}}\label{gen-trans-t}
        \\&+\frac{\eta^{(n)}}{2E_{\text{Pl}}}\left(t\sum_{r=0}^n \alpha_r E^{n-r}|p|^r+x\sum_{r=0}^n \beta_r E^{n-r}|p|^r\right)\,,\nonumber\\
        x' = &\,\,\, (tv +x)\gamma +\frac{\eta^{(n)}}{2} t v\gamma \left(\frac{E}{E_{\text{Pl}}}\right)^n\left(\gamma ^2-1\right)^{\frac{n+2}{2}} \\&+\frac{\eta^{(n)}}{2E_{\text{Pl}}}\left(x\sum_{r=0}^n \delta_r E^{n-r}|p|^r+t\sum_{r=0}^n \lambda_r E^{n-r}|p|^r\right)\,,\nonumber
    \end{align}
where we isolated the terms from the sum which are non-vanishing in the rest frame, i,e,\ $p=0=x$, $E=m$, for clarity.

Imposing the isometry condition $F^2(\dot{t}',\dot{x}')=F^2(\dot{t},\dot{x})$ on this transformation and noticing that $E=m\dot{t}/\sqrt{\dot{t}^2-\dot{x}^2}+{\cal O}(m/E_{\text{Pl}})$ and $p=-m\dot{x}/\sqrt{\dot{t}^2-\dot{x}^2}+{\cal O}(m/E_{\text{Pl}})$ are conserved functions of velocities, we derive an expression involving powers and factors of $\dot{t}$ and $|\dot{x}|$, besides terms like $(|\dot{x}|+v\dot{t})^n$, for which we can use the binomial theorem to express it in terms of combinations of $\dot{t}^{n-r}|\dot{x}|^r$. Furthermore, imposing that this transformation should reduce to that of Eqs.\eqref{transf1}, \eqref{transf2} when $p=0=x$ and $E=m$, we find the following conditions
\begin{align}
    \alpha_0&=0=\lambda_0,\, & \beta_0&=-v^{1+n}\gamma^{3+n},\,\\ 
    \delta_0&=-v^n\gamma^{1+n}(\gamma^2-1),\\
    \alpha_n&=v\beta_n+\gamma-\gamma^{-1},\,& \lambda_n&=\beta_n+v\gamma(1+\gamma^n)\,,\\
    \delta_n&=v\beta_n+\gamma^{-1}+\gamma^{1+n}\,,
\end{align}
and for $1\leq r\leq n-1$,
\begin{align}
    \alpha_r&=v\beta_r,\,\qquad  \lambda_r=\beta_r+v^{1+n-r}\gamma^{1+n}\binom{n}{r}\,,\\
    \delta_r&=v\beta_r+v^{n-r}\gamma^{1+n}\binom{n}{r}\,,\label{last-gen-cond}
\end{align}
where $\binom{n}{r}$ is the binomial coefficient. We see that we have the freedom to choose $n$ arbitrary functions of $v$, namely $\beta_r\, (1\leq r\leq n)$ that should be null when $v\rightarrow 0$ in order to guarantee that we recover the identity transformation when the velocity is zero. As this transformation preserves the Finsler function, it also preserves the metric and consequently the MDR \eqref{mdr} that is calculated from the norm of the momenta \cite{Girelli:2006fw,Barcaroli:2015xda,Lobo:2016xzq}.

A simple choice for the $\beta_r$ is setting all $\beta_r=0$ for $1\leq r\leq n$. Another possible choice is fixing $\beta_r$ is the comparison of this transformation with one arising from the action of boosts in a quantum algebraic approach. For example, similar ambiguities are found at the bicrossproduct basis of $\kappa$-Poincar\'e-inspired Finsler isometries, which could be fixed by comparing the generators of the transformations found from the geometric and algebraic approaches \cite{Amelino-Camelia:2014rga}.

This set of deformed relativistic transformations constitute an isometry of the Finsler function that preserves the spacetime origin. Other isometries are rotations (which should be unmodified within this isotropic approach) and translations. The infinitesimal isometries for the $n=1$ case have been calculated in \cite{Amelino-Camelia:2014rga} where, in particular, deformed translations are found. These transformations play a fundamental role in avoiding unphysical nonlocalities in the DSR scenario, as  pointed out in \cite{Hossenfelder:2010tm}, and are related to the concept of relative locality \cite{Amelino-Camelia:2011lvm,Amelino-Camelia:2011uwb}, which shows that ambiguities in the location of events in the DSR scenario is a coordinate artifact. Therefore, for consistency, we  consider that the set of symmetries connecting observers within this approach is given by the isometries of the Finsler function.

\section{Conclusion}
Phenomenological models which break Lorentz invariance lead to a modified Lorentz factor which encodes the time dilation between different observer frames. This prediction triggered the search for such phenomenology using the mass content of EAS from UHECR data \cite{Trimarelli:2022wdd}, thus using deformations of particle lifetimes as a window to Planck scale physics.
\par
In this letter, we have used Finsler geometry to show that actually a similar correction (with the opposite sign) emerges naturally from an approach that {\it deforms}, rather than breaks, Lorentz symmetry, such that the modified dispersion relation's form is kept invariant when transforming between frames in a way that not only preserves the relativity principle, but also the so-called clock postulate (the observer's proper time is the line element of its trajectory) from special relativity to a deformed special relativity. This can be seen in the discussion that leads to Eqs.\eqref{cp-lifetime} and \eqref{cp-dsr}. Therefore, unintentionally, what has been considered in previous analysis using UHECR data would be a {\it deformation} instead of a {\it violation} of Lorentz symmetry with basically opposite signs on the correction. The phenomenological consequences of these two scenarios are manifestly different. For instance, some processes that are forbidden in the Lorentz-invariance case would be allowed in the LIV scenario but not in DSR. These additional observables would ultimately allow us to distinguish between these scenarios.
\par
Besides that, we have shown that one must not use a velocity given by $\beta_{\text{LIV}}=|p|/m\, \gamma_{\text{LIV}}$ (Eq.\eqref{vliv}), as done in \cite{Trimarelli:2022wdd} since it is incompatible with the actual velocity of a particle in the lab frame, which must be read from the dispersion relation, whose expression can also be naturally derived from our analysis, as can be seen in the discussion that surrounds Eqs.\eqref{vdsr} and \eqref{v-vdsr}.
\par
We also generalized this result to a transformation between two lab frames that move relative to each other with velocity $v$, given by Eqs.\eqref{gen-trans-t}-\eqref{last-gen-cond}, and that reduces to the previous case in the comoving limit $x=0=p$ and $E=m$. We believe that the search for quantum gravity effects from cosmic-ray data could benefit from the findings of this letter within the scope of the deformation of Lorentz symmetry and a next natural step would be to consider these findings in future analyses. As a final remark, as UHECR observatories improve their detection techniques and capabilities of reconstruction of air showers, the prospects for detecting the effects discussed here will become even better, especially with future facilities~\cite{Coleman:2022abf}.

\section{Acknowledgments}
P. H. M. thanks Coordena\c c\~ao de Aperfei\c coamento de Pessoal de N\'ivel Superior - Brazil (CAPES) - Finance Code 001 for financial support. I. P. L. was supported by the National Council for Scientific and Technological Development - CNPq grant 306414/2020-1 and by the grant 3197/2021, Para\'iba State Research Foundation (FAPESQ). C. P. is funded by the excellence cluster QuantumFrontiers funded by the Deutsche Forschungsgemeinschaft (DFG, German Research Foundation) under Germany’s Excellence Strategy – EXC-2123 QuantumFrontiers – 390837967. R. A. B. is funded by the ``la Caixa'' Foundation (ID 100010434) and the European Union's Horizon~2020 research and innovation program under the Marie Sklodowska-Curie grant agreement No 847648, fellowship code LCF/BQ/PI21/11830030. V. B. B. was supported by the National Council for Scientific and Technological Development - CNPq grant 307211/2020-7. The authors like to acknowledge networking support by the COST Action CA18108.
\bibliographystyle{elsarticle-num} 
\bibliography{PLB_version}






\end{document}